\documentclass[twocolumn,showpacs,preprintnumbers,amsmath,amssymb,nofootinbib]{revtex4-1}

\usepackage{graphicx}
\usepackage{dcolumn}
\usepackage{bm}
\usepackage{color}

\def\bequ{\begin{equation}}
\def\eequ{\end{equation}}
\def\be{\begin{equation}}
\def\ee{\end{equation}}

\begin{document}


\title{Quasinormal modes and late time tails of perturbation fields on a Schwarzschild--like black hole with a global monopole in the Einstein--bumblebee theory}

\author{Xiaolin Zhang}
\author{Mengjie Wang}
\email{Corresponding author: mjwang@hunnu.edu.cn}
 \author{Jiliang Jing}
\affiliation{\vspace{2mm}
Department of Physics, 
Synergetic Innovation Center for Quantum Effects and Applications, 
and Institute of Interdisciplinary Studies, 
Hunan Normal University, Changsha, Hunan 410081, P. R. China 
\vspace{1mm}}

\date{\today}%
\begin{abstract}
In this paper we complete a systematic study on quasinormal modes (QNMs) and late time tails for scalar, Dirac and Maxwell fields on a spherically symmetric Schwarzschild--like black hole with a global monopole in the Einstein--bumblebee theory. To look for QNMs, we solve the equations of motion for all perturbation fields considered herein numerically, by employing both the matrix and the WKB methods, and find good agreements for numeric data obtained by these two techniques in the regime when both are valid. The impact of the bumblebee parameter $c$, the monopole parameter $\eta^2$ and the multipole number $\ell$ on the fundamental quasinormal frequency is analyzed in detail. Our results are shown in terms of the quasinormal frequency measured by $\sqrt{1+c}\,M$, where $M$ is a black hole mass parameter. We observe, by increasing the parameter $c$ ($\eta^2$) with fixed first few $\ell$, that the real part of QNMs increases for all spin fields; while the magnitude of the imaginary part decreases for scalar and Dirac fields but increases for Maxwell fields. By increasing the multipole number $\ell$ with fixed other parameters, we disclose that the real part of QNMs for all spin fields increases while the magnitude of the imaginary part decreases for scalar and Dirac fields but increases for Maxwell fields. In the eikonal limit ($\ell\gg n$), QNMs for all spin fields coincide with each other and the real part scale linearly with $\ell$. In particular, the asymptotic QNMs approach the corresponding results given by the first order WKB formula, and \textit{only} the real part of QNMs is dependent on the bumblebee and monopole parameters. In addition, it is shown that the late time behavior is determined not only by the multipole number but also by the bumblebee and monopole parameters, and is distinct for bosonic and fermonic fields. Moreover, the presence of the bumblebee (monopole) field makes the spin fields decay faster. Our results indicate, both in the context of QNMs and late time tails, that the bumblebee field and the monopole field play the same role in determining the dynamic evolution of perturbation fields.
\end{abstract}
\maketitle

\section{Introduction}
As the standard model of gravity, General Relativity (GR) has been extensively studied for decades. Although remarkable achievements have been made, the discussion on the extension of GR, see for example~\cite{Sotiriou:2008rp,DeFelice:2010aj,Nojiri:2010wj,Capozziello:2011et,Clifton:2011jh} and references therein, has never been ceased. This is motivated, on the theoretic side, to construct renormalizable theories of gravity; while on the phenomenological side, to explain the origin of dark energy and dark matter.

Among numerous modified theories of gravity, one possible extension of GR is the gravitational theory with Lorentz symmetry violation~\cite{Jacobson:2000xp,Jacobson:2007veq,Horava:2009uw,Jacobson:2007veq}. Lorentz symmetry is a pillar of modern physics while it may \textit{not} be valid at the Planck scale~\cite{Mattingly:2005re,Jacobson:2005bg}. It is thus interesting to test possible Lorentz symmetry violation signatures in the gravitational sector. As one of the Lorentz violation gravitational theories, the bumblebee gravity has been received much attention recently. In such a model, the Lorentz symmetry is broken spontaneously, which is trigged by a vector field known as the bumblebee field~\cite{Kostelecky:1989jw,Kostelecky:2000mm,Kostelecky:2003fs}.

In order to test the Lorentz symmetry in the Einstein--bumblebee theory, one needs exact solutions. A static spherically symmetric neutral black hole (BH) solution, only having a nonvanishing radial component of the bumblebee field, was first obtained in~\cite{Casana:2017jkc}. Later on, static spherically symmetric BH solutions were given, in four dimensions with a cosmological constant~\cite{Maluf:2020kgf}, with a global monopole~\cite{Gullu:2020qzu}, with a Gauss-Bonnet term~\cite{Ding:2021iwv}, with a nonvanishing temporal component of the bumblebee field~\cite{Xu:2022frb}, and in higher~\cite{Ding:2022qcy} and lower dimensions~\cite{Ding:2023niy}. By adding a rotation, a four dimensional stationary axisymmetric BH solution was also found~\cite{Ding:2020kfr}, in the slow rotation approximation.

Given the above BH solutions, the Lorentz symmetry breaking effects have been explored in various aspects, including, for example, BH thermodynamics~\cite{Gomes:2018oyd,Kanzi:2019gtu,Karmakar:2023mhs}, geodesics and applications to gravitational lensing and BH shadow~\cite{Li:2020wvn,KumarJha:2020ivj,Ovgun:2018ran}, quasinormal modes (QNMs)~\cite{Oliveira:2018oha,Oliveira:2021abg,Chen:2023cjd,Liu:2022dcn,Gogoi:2022wyv}, and gravitational waves~\cite{Liang:2022gdk,Liang:2022hxd,Amarilo:2019lfq}. In particular, with the discovery of gravitational waves~\cite{LIGOScientific:2016aoc,Jing:2023vzq,Jing:2022vks,Jing:2021ahx,Zou:2021lkj,He:2019oun} and the observation of BH shadow~\cite{EventHorizonTelescope:2019dse,Chen:2022scf,Liu:2022ruc,Zhou:2021cef}, more gravitational tests may be conducted in the future. In this paper, we perform a complete and systematic study on the dynamic evolution of various perturbation fields ($i.e.$ scalar, Dirac and Maxwell fields) on a Schwarzschild--like BH with a global monopole in the Einstein-Bumblebee theory.

The dynamic evolution of arbitrary spin fields around BHs consists of three stages: the initial wave burst, the damped oscillations (also well known as QNMs), and the power law late time tails~\cite{Nollert:1999ji,Kokkotas:1999bd,Berti:2009kk,Konoplya:2011qq}. Among them, QNMs and late time tails are significant research subjects in BH physics since, for the former case QNMs have various applications ranging from gravitational waves to high energy physics; while for the latter case the late time tails are closely related to some fundamental properties of BHs, such as the no hair conjecture. 

In the Einstein--bumblebee theory, an exact spherically symmetric BH solution sourced by a global monopole is presented in~\cite{Gullu:2020qzu}. The global monopoles are one of topological defects, which may be formed in the early universe by spontaneously breaking the global $O(3)$ symmetry to $U(1)$~\cite{Kibble:1976sj,Vilenkin:1984ib}. Now that both the bumblebee field and the global monopole are related to spontaneous symmetry breaking, albeit corresponding to different symmetries, one may wonder if both have similar physical consequences. We identify, in the context of QNMs and late time tails, that this is \textit{indeed} the case. 

The structure of this paper is organized as follows. In Section~\ref{seceq} we introduce the background geometry and present equations of motion for scalar, Dirac and Maxwell fields. In Section~\ref{secqnm} we describe a matrix method and the WKB method to look for QNMs of perturbation fields we considered herein numerically and illustrate the corresponding numeric results. In Section~\ref{secltt} we explore late time behaviors of various spin fields and close with final remarks in the last section. To further clarify our results, we rewrite equations of motion for scalar, Dirac and Maxwell fields in terms of the frequency $\tilde{\omega}$ in the Appendix.

\section{background geometry and field equations}
\label{seceq}
In this section, we briefly describe the background geometry, and derive the corresponding equations of motion for scalar, Dirac and Maxwell fields.

For our interests herein, we consider a Schwarzschild--like spherically symmetric BH with a global monopole in the Einstein-bumblebee gravity~\cite{Gullu:2020qzu}
\begin{equation}
ds^{2}=-f(r)d t^{2}+\dfrac{1+c}{f(r)}d r^{2}+\bar{\eta}^{2} r^{2}\left(d \theta^{2}+\sin ^{2} \theta d \varphi^{2}\right)\;,\label{metric}
\end{equation}
with
\begin{equation}
f(r)=1-\frac{2 M}{r}\;,\label{metricfunc}
\end{equation}
where the bumblebee parameter $c=\xi b^{2}$, $b^2$ is a real positive constant and $\xi$ is the nonminimal coupling constant of gravity-bumblebee interaction, and $\bar{\eta}^2$ is related to the monopole $\eta^{2}$ through $\bar{\eta}^2=1-\eta^{2}$. The black hole event horizon $r_+$ is determined by $f(r_+)=0$, and the Hawking temperature is given by~\cite{Gullu:2020qzu}
\begin{equation}
T_H=\dfrac{1}{4\pi r_+\sqrt{1+c}}\;.\label{hawkingT}
\end{equation}

The dynamic properties of the above background given in Eq.~\eqref{metric} may be explored by introducing perturbation fields. Here, we perform a systematic study of various spin test fields, by deriving equations of motion for massless scalar, Dirac and Maxwell fields. Note that the following equations of motion are derived under the assumption of the time dependence $e^{-i\omega t}$.

For a massless scalar field, it is governed by the Klein-Gordon equation 
\begin{equation}
\dfrac{1}{\sqrt{-g}}\partial_{\mu}\left(\sqrt{-g} g^{\mu \nu}\partial_{\nu} \Phi\right)=0\;,
\end{equation}
which, by expanding the scalar field in terms of the scalar spherical harmonics, turns into
 \begin{equation}
\left(\dfrac{d^2}{dr_*^2}+\omega^2-V_{s}(r)\right)R(r)=0\;,\label{eqS}
\end{equation}
where $\omega$ is the frequency and $R(r)$ is the radial part of the scalar field $\Phi$. Here the tortoise coordinate and the effective potential are given by
\begin{equation}
\dfrac{dr_*}{dr}=\dfrac{\sqrt{1+c}}{f(r)}\;,\label{tortoiseeq}
\end{equation}
and
\begin{equation}
V_s(r)=f(r)\left(\frac{2M}{(1+c)r^3}+\frac{\lambda}{\bar{\eta}^2r^2}\right)\;,\label{potentialS}
\end{equation}
where $\lambda$ is related to the multipole number $\ell$ by $\lambda=\ell(\ell+1)$.

For a massless Dirac field, it obeys the Dirac equation 
\begin{equation}
\gamma^{\alpha}(\partial_{\alpha}-\Gamma_{\alpha})\Psi=0\;,\label{eqD1}
\end{equation}
where the matrices $\gamma^{\alpha}$ may be constructed by~\cite{Unruh:1973bda,Wang:2017fie,Wang:2019qja}
\begin{equation}
\begin{array}{c}
\gamma ^{t} = \dfrac{1}{\sqrt{f(r)}} \gamma ^{0}\;,\;\;\;\quad \gamma ^{r} = \sqrt{\dfrac{f(r)}{1+c} } \gamma ^{3}\;,\quad \\ \gamma ^{\theta} = \dfrac{1}{\bar{\eta}r} \gamma ^{1}\;,\;\;\;\;\;\;\;\;\quad \gamma ^{\varphi} = \dfrac{1}{\bar{\eta}r \sin \theta} \gamma ^{2}\;,
\end{array}
\end{equation}
with the ordinary flat spacetime Dirac matrices $\gamma^j$ ($j=0,1,2,3$) in the Bjorken--Drell representation~\cite{Bjorken:100769}, and the spin connection is defined as
\begin{equation}
\Gamma_{\alpha}=-\dfrac{1}{8}\left(\gamma^{a} \gamma^{b}-\gamma^ b\gamma ^a\right) \Sigma_{a b \alpha}\;,
\end{equation}
with
\begin{equation}
\Sigma_{a b \alpha}=e_{a}^{\beta}(\partial_{\alpha} e_{b \beta}-\Gamma_{\beta \alpha}^{\rho} e_{b \rho})\;.
\end{equation}
Following the procedures presented in~\cite{Unruh:1973bda,Wang:2017fie,Wang:2019qja}, one may get the radial part of the Dirac equation
\begin{equation}
\left(\frac{d^2}{dr_{*}^2}+\omega^2-V_{d}(r)\right)R(r)=0\;,\label{eqD2}
\end{equation}
where the effective potential of Dirac fields is given by
\begin{align}
V_d(r)=&\dfrac{\lambda^{2}f(r)}{\bar{\eta}^2r^2}-\dfrac{rf(r)}{4(1+c)}\left(\dfrac{f'(r)}{r}\right)^\prime+\dfrac{f'(r)^2}{16(1+c)}\nonumber\\&-\dfrac{f(r)^2}{4(1+c)r^2}\pm\dfrac{i\omega f(r)}{\sqrt{1+c}\,r}\mp\dfrac{i\omega f'(r)}{2\sqrt{1+c}}\;,\label{potentialD_complex}
\end{align}
with $\lambda^2=(\ell+\tfrac{1}{2})^2$~\cite{Wang:2017fie,Wang:2019qja}. Here $^\prime$ denotes the derivative with respect to $r$, and $\pm$ describes two different chiralities of Dirac fields. 

As one may observe, from Eq.~\eqref{potentialD_complex}, that the effective potential of Dirac fields is \textit{complex}. By using the transformation theory or by following the procedures provided in~\cite{Chandrasekhar:1985kt}, one may obtain the Schrodinger--like equation for Dirac fields with the \textit{real} effective potential, which is
\begin{equation}
V_d(r)=\frac{f(r) \kappa^2}{\bar{\eta}^2 r^2}\pm\frac{\kappa \sqrt{f(r)}}{\sqrt{1+c}\,\bar{\eta}\, r^2}\left[\frac{f^{\prime}(r) r}{2}-f(r)\right]\;,\label{potentialD_real}
\end{equation}
where, again, $\pm$ represents two different chiralities of Dirac fields, and $\kappa=\ell+1$ ($\kappa=\ell$) for $+$ ($-$) sign in the above equation.

As we have checked, QNMs of Dirac fields with different chiralities are the same, so that one may take either of them ($i.e.$ $-$ sign or $+$ sign in effective potentials) to perform numeric calculations.

For a Maxwell field, it is followed by the Maxwell equation
\begin{equation}
\dfrac{1}{\sqrt{-g}}\partial_{\nu}\left(\sqrt{-g} g^{\mu \alpha}g^{\nu \beta}F_{\alpha\beta} \right )=0\;,\label{eqM1}
\end{equation}
where the Maxwell tensor is $F_{\mu\nu}=\partial_{\mu}A_{\nu}-\partial_{\nu}A_{\mu}$ and $A_{\mu}$ is a four dimensional vector potential. In the Regge--Wheeler--Zerilli formalism~\cite{Regge:1957td,Zerilli:1970se}, one may decompose $A_{\mu}$ in terms of the scalar ($Y_{\ell m}$) and vector ($\boldsymbol{S}_{\ell m}$ and $\boldsymbol{Y}_{\ell m}$) spherical harmonics~\cite{Ruffini:1973}
\begin{equation}
A_{\mu}=\sum_{\ell, m}e^{-i \omega t}\left(\left[\begin{array}{c} 0 \\
0 \\
a^{\ell m}(r) \boldsymbol {S}_{\ell m}\end{array}\right]+\left[
\begin{array}{c}j^{\ell m}(r)Y_{\ell m}  \\
h^{\ell m}(r)Y_{\ell m} \\
k^{\ell m}(r)\boldsymbol {Y}_{\ell m}  
\end{array}\right]\right)\;,\label{Vpotential}
\end{equation}
where the explicit form of the vector spherical harmonics is given in~\cite{Wang:2021upj,Wang:2021uix,Lei:2021kqv}, and $m$ is the azimuthal number. Note that the first/second term in the right hand side of Eq.~\eqref{Vpotential} has parity $(-1)^{\ell+1}$/$(-1)^\ell$ so that we dub the former/latter the axial/polar modes. Substituting Eq.~\eqref{Vpotential} into Eq.~\eqref{eqM1}, one obtains the radial part of the Maxwell equation
\begin{equation}
\left(\frac{d^2}{dr_{*}^2}+\omega^2-V_{m}(r)\right)R(r)=0\;,\label{eqM2}
\end{equation}
where the effective potential of the Maxwell field is given by 
\begin{equation}
V_{m}(r)=f(r)\frac{\lambda}{\bar{\eta}^2r^2}\;,\label{potentialM}
\end{equation}
with $\lambda=\ell(\ell+1)$. Notice that Eq.~\eqref{eqM2} is held not only for axial but also for polar modes.

To look for QNMs, one has to assign an ingoing wave boundary condition at the event horizon and an outgoing wave boundary condition at infinity to all spin fields we studied in this paper. The former condition at the event horizon leads to
\begin{equation}
R\sim e^{-i\omega r_*}\;,\label{ingoingBC}
\end{equation}
while the latter condition at infinity gives
\begin{equation}
R\sim e^{i\omega r_*}\;.\label{outgoingBC}
\end{equation}
Note that these boundary conditions are only held for perturbation equations with real effective potentials. 

For Dirac fields with the complex potential given in Eq.~\eqref{potentialD_complex}, the ingoing wave boundary condition at the event horizon should be altered into the form
\begin{equation}
R\sim e^{-i\bar{\omega} r_*}\;,\label{ingoingBC_Dirac_complex}
\end{equation}
where 
\begin{equation}
\bar{\omega}\equiv\omega+\dfrac{i}{4\sqrt{1+c}\;r_+}\;,\nonumber
\end{equation}
while the boundary condition at infinity keep the same form as given by Eq.~\eqref{outgoingBC}.

\section{quasinormal modes}
\label{secqnm}
With equations of motion of scalar, Dirac and Maxwell fields and the corresponding physically relevant boundary conditions at hand, we are ready to compute QNMs. To achieve this goal, numeric methods have to be applied. Here we introduce two methods employed in our paper, $i.e.$ the WKB and the matrix approaches.
\subsection{Methods}
\subsubsection{The WKB method}
The WKB method, proposed initially by Schutz and Will~\cite{Schutz:1985km} and based on the JWKB approximation~\cite{jeffreys1925certain,Wentzel:1926aor,Kramers:1926njj,brillouin1926mecanique}, is a powerful semianalytic technique to calculate QNMs. This method is typically applied to the Schrodinger--like equation with real effective potentials which have a potential barrier and tend to constant values at the event horizon and at infinity. The basic idea of this method is to match the asymptotic WKB solutions at the two boundaries (the event horizon and infinity) with the Taylor expansion near the potential peak through the two turning points.

The WKB approach, at the first order, leads to the following implicit eigenvalue formula labeled by an overtone number $n$~\cite{Schutz:1985km}
\begin{equation}
i\dfrac{Q_0}{\sqrt{2Q_0^{\prime\prime}}}=n+\frac{1}{2}\;,\;\;\;\;\;\;n=0, 1, 2, \cdots\;,\label{WKBorder1}
\end{equation}
where $Q_0$ is the maximum value of the function $Q$ with $Q\equiv\omega^2-V$, and $^{\prime\prime}$ denotes the second derivative of a function with respect to the tortoise coordinate $r_*$. Here for scalar fields $V=V_s$ given in Eq.~\eqref{potentialS}, for Dirac fields $V=V_d$ given in Eq.~\eqref{potentialD_real}, and for Maxwell fields $V=V_m$ given in Eq.~\eqref{potentialM}. This method was later extended to higher orders~\cite{Iyer:1986np,Konoplya:2003ii,Matyjasek:2017psv,Konoplya:2019hlu}. In this paper, we employ the six order WKB formula
\begin{equation}
i\dfrac{Q_0}{\sqrt{2Q_0^{\prime\prime}}}-\sum_{j=2}^{6} \Lambda_{j}=n+\frac{1}{2}\;,\label{WKBorder6}
\end{equation}
where $\Lambda_{j}$ are higher order correction terms  and their explicit forms may be found in~\cite{Iyer:1986np,Konoplya:2003ii}, to calculate QNMs for various spin fields. Neglecting these correction terms, the first order WKB formula~\eqref{WKBorder1} is recovered.

One shall notice that the accuracy of the WKB method depends closely on the relation between the multipole number $\ell$ and the overtone number $n$. More specifically, this method is valid in the regime $\ell\gg n$, and therefore it is more appropriate to search for low-lying modes. Beyond the above mentioned regime, one has to employ the other methods to figure out QNMs. In the following we introduce one of such methods, $i.e.$ the matrix method.

\subsubsection{The matrix method}
The matrix method is another efficient approach to solve QNMs~\cite{Lin:2016sch,Lin:2017oag}. The basic idea of this method is as follows: ($i$) introducing equally spaced grid points in the internal $[0,1]$, ($ii$) expanding an unknown function around each grid points by the Taylor series, ($iii$) imposing the proper boundary conditions and then transforming the perturbation equation into the matrix form. 

More specifically, we first transform the radial function from $R(r)$ to $\chi(r)$ through the relations
\begin{equation}
R(r)=e^{i\omega\sqrt{1+c}\,r}r^{2i\omega\sqrt{1+c}\,r_+}(r-r_+)^{-i\omega\sqrt{1+c}\,r_+}\chi(r)\;,\nonumber
\end{equation}
for scalar and Maxwell fields by considering the boundary conditions given in Eqs.~\eqref{ingoingBC} and~\eqref{outgoingBC}, and
\begin{equation}
R(r)=e^{i\omega\sqrt{1+c}\,r}r^{i(\omega+\bar{\omega})\sqrt{1+c}\,r_+}(r-r_+)^{-i\bar{\omega}\sqrt{1+c}\,r_+}\chi(r)\;,\nonumber
\end{equation}
for Dirac fields~\footnote{In the context of the matrix method, for numeric accuracy and efficiency we usually solve perturbation equations with rational potential. We therefore implement the complex potential~\eqref{potentialD_complex} to look for Dirac QNMs.} by considering the boundary conditions given in Eqs.~\eqref{ingoingBC_Dirac_complex} and~\eqref{outgoingBC}, such that the boundary conditions for $\chi$ are satisfied automatically.

We then make the coordinate and functional transformations
\begin{equation}
x=1-\dfrac{r_+}{r}\;,\;\;\;\;\;\;\;\;\xi(x)=x(1-x)\chi(x)\;,\nonumber
\end{equation}
so that the perturbation equations and the corresponding boundary conditions may be rewritten as
\begin{align}
&\mathcal{T}_2(x)\xi''(x)+\mathcal{T}_1(x,\omega)\xi'(x)+\mathcal{T}_0(x,\omega)\xi(x)=0\;,\nonumber\\
&\xi(0)=\xi(1)=0\;,\label{EqMatrix}
\end{align}
where $^\prime$ denotes the derivative with respect to $x$, and the functions $\mathcal{T}_j (j=0,1,2)$ can be derived straightforwardly. 

Finally, by discretizing Eq.~\eqref{EqMatrix} on the equally spaced grid points and by using the corresponding differential matrices~\cite{Lin:2016sch,Lin:2017oag}, Eq.~\eqref{EqMatrix} turns into an algebraic equation with the matrix form
\begin{equation}
\mathcal{M}(\omega)\xi=0\;,
\end{equation}
which leads to quasinormal frequencies by imposing the condition $|\mathcal{M}(\omega)|=0$, and where $|\mathcal{M}(\omega)|$ is the determinant of the matrix $\mathcal{M}(\omega)$. 

\subsection{Results}
With the above numeric methods at hand, we are ready to calculate QNMs. Before we present numeric results, a few remarks are in order. Considering the limitation of the WKB method, numeric data presented in this part are normally generated by the matrix method, and they are also checked by the sixth order WKB approach when it is applicable. In the regime when both methods are valid, we observe that the data obtained by these two approaches agree well, which indicates the correctness of our results.

In the numeric calculations, we work in the unit of $M=1$. In order to identify and understand the impact of the bumblebee field parameter $c$ and the monopole parameter $\eta^2$ on quasinormal frequencies, we present our results in terms of $\tilde{\omega}$ instead of $\omega$, with the definition $\tilde{\omega}=\sqrt{1+c}\,\omega$. This implies that $\tilde{\omega}$ is the quasinormal frequency measured in terms of $\sqrt{1+c}M$. As one may observe in the following, by using $\tilde{\omega}$, the effects of the bumblebee parameter $c$ and the monopole parameter $\eta^2$ on QNMs are the same. And this may be easily understood based on the fact, as shown in the Appendix~\ref{app}, that both the bumblebee parameter and the monopole parameter determine the quasinormal frequency $\tilde{\omega}$ through changing the multipole number $\ell$.
 
\begin{figure*}
\begin{center}
\includegraphics[clip=false,width=0.5\textwidth]{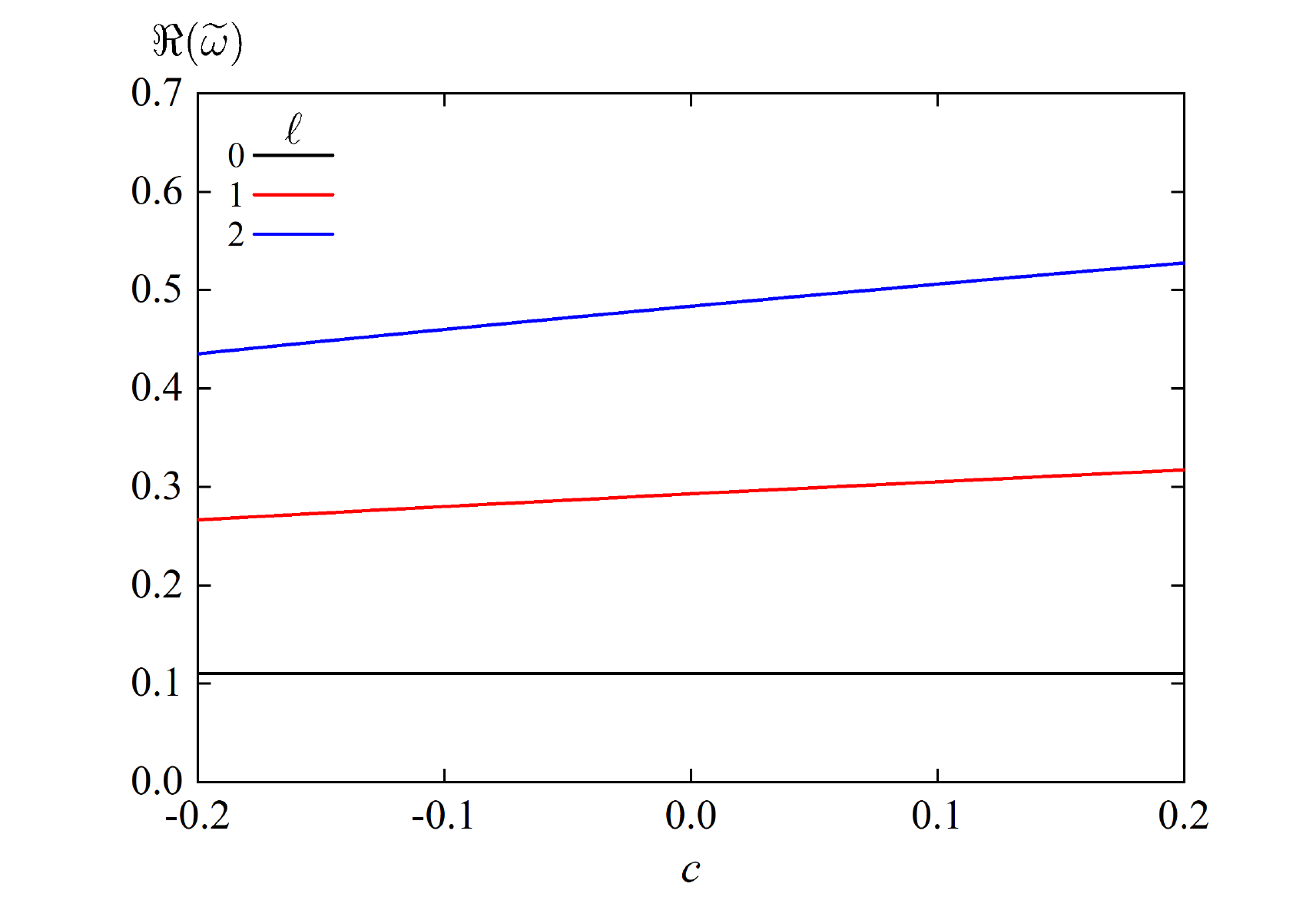}\;\;\includegraphics[clip=false,width=0.5\textwidth]{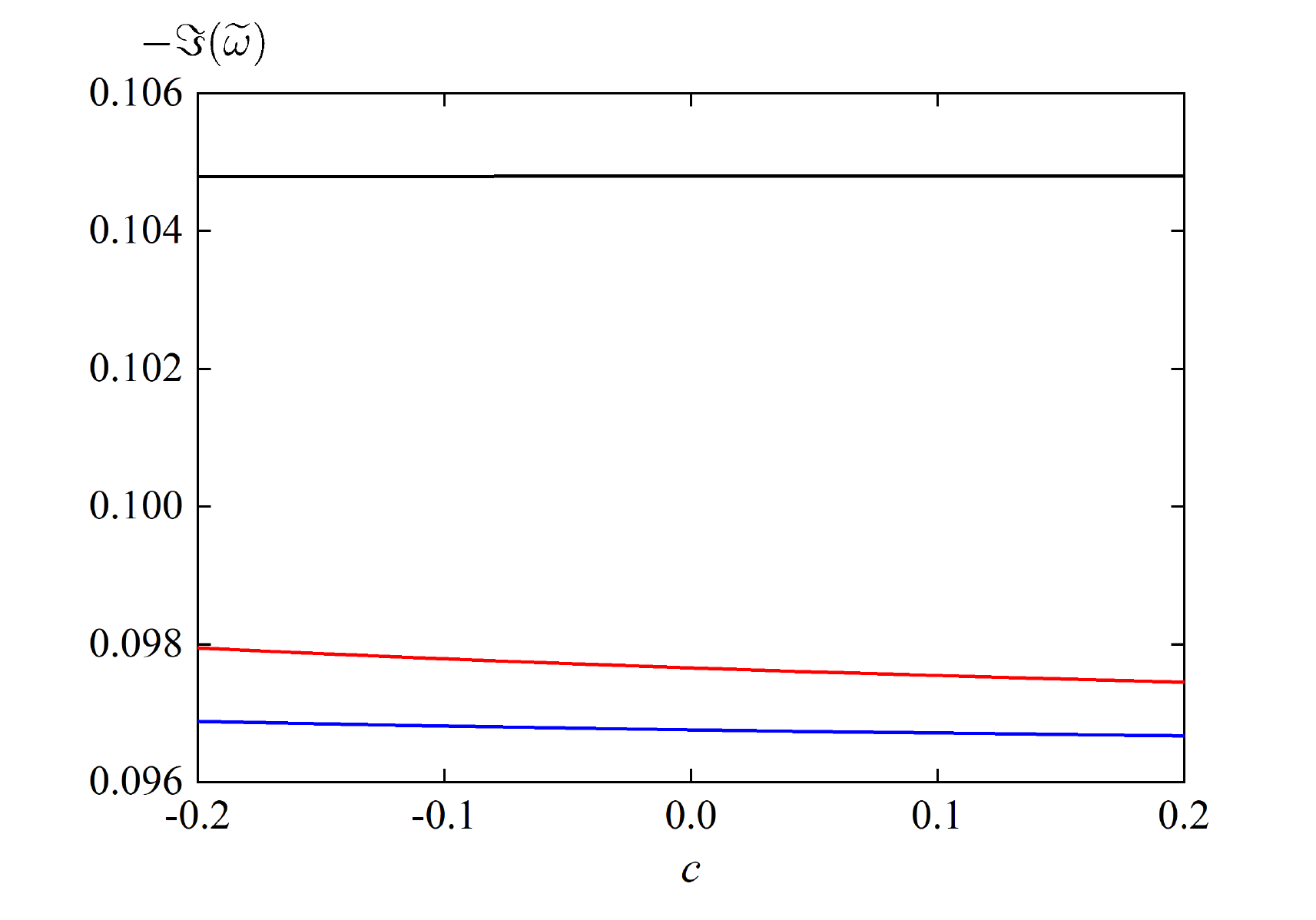}
\end{center}
\caption{\label{scalarQNMs_Bumblebee} (color online). Real (left) and imaginary (right) parts of fundamental QNMs for scalar fields in terms of the bumblebee parameter $c$, for $\ell=0$ (black), $\ell=1$ (red) and $\ell=2$ (blue), with fixed $\eta^2=0$.}
\end{figure*}

\begin{figure*}
\begin{center}
\includegraphics[clip=false,width=0.5\textwidth]{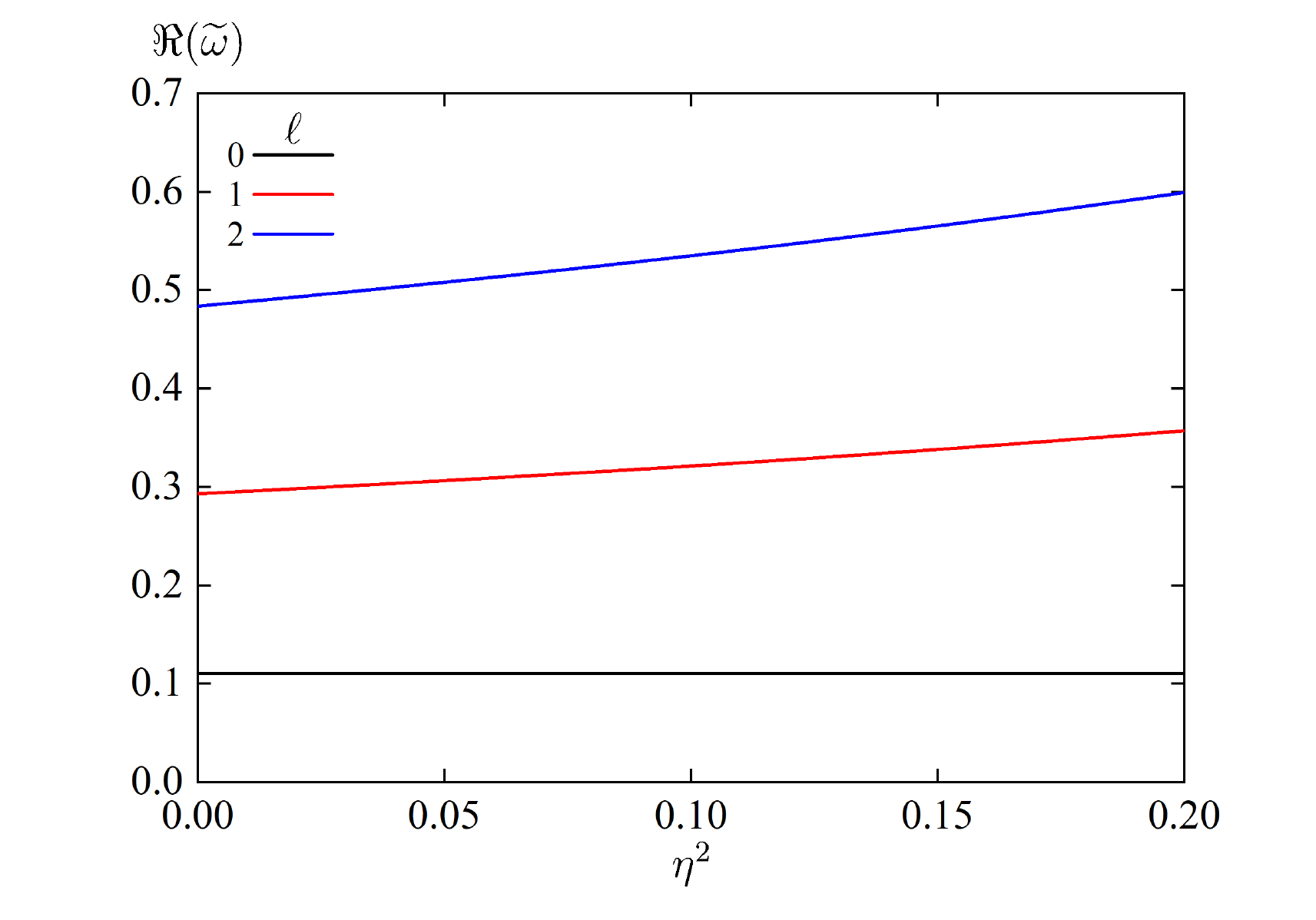}\;\;\includegraphics[clip=false,width=0.5\textwidth]{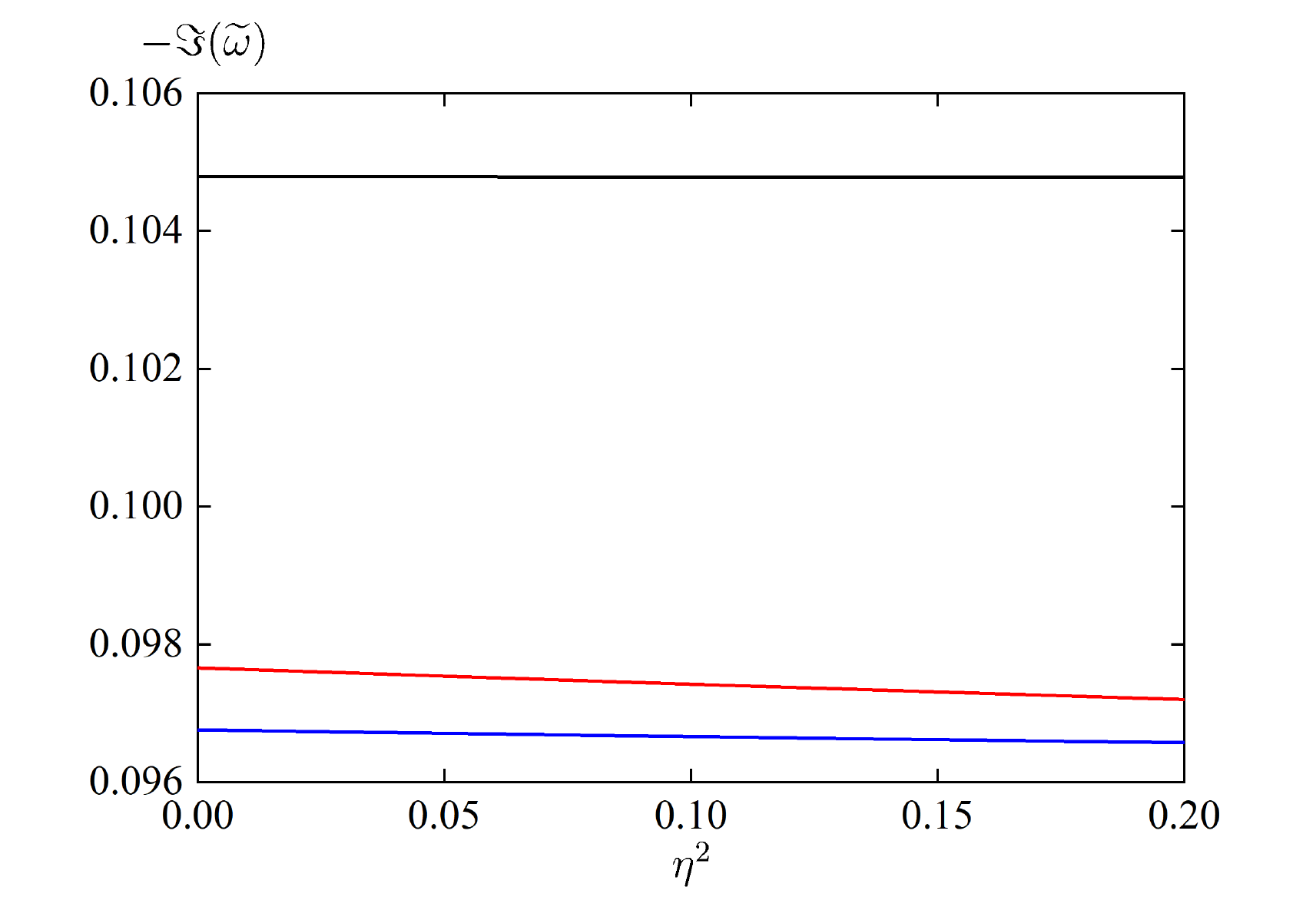}
\end{center}
\caption{\label{scalarQNMs_monopole} (color online). Real (left) and imaginary (right) parts of fundamental QNMs for scalar fields in terms of the monopole parameter $\eta^2$, for $\ell=0$ (black), $\ell=1$ (red) and $\ell=2$ (blue), with fixed $c=0$.}
\end{figure*}

We start by displaying numeric results of scalar fields. In Fig.~\ref{scalarQNMs_Bumblebee}, we explore fundamental scalar QNMs by varying the bumblebee parameter $c$, with the monopole parameter $\eta^2=0$ and for the multipole number ranging from $\ell=0$ (black), $\ell=1$ (red) to $\ell=2$ (blue). It is shown that, for the $\ell=0$ mode, the bumblebee parameter \textit{does not} affect quasinormal frequencies~\footnote{This may be easily understood because, based on our analysis in the Appendix~\ref{app}, $\ell=0$ leads to $\tilde{\ell}=0$ so that the bumblebee parameter is not shown up anymore in the equations of motion.}; while for $\ell\neq0$ modes, as $c$ increases, the real part of quasinormal frequency increases but the magnitude of the imaginary part decreases. The impact of the monopole parameter $\eta^2$ on scalar QNMs, as shown in Fig.~\ref{scalarQNMs_monopole}, is similar to the bumblebee parameter. This fact, held not only for scalar but also for Dirac and Maxwell fields as we will see later, implies that the bumblebee parameter plays the same role as the monopole parameter, in determining QNMs. 

\begin{figure*}
\begin{center}
\includegraphics[clip=false,width=0.5\textwidth]{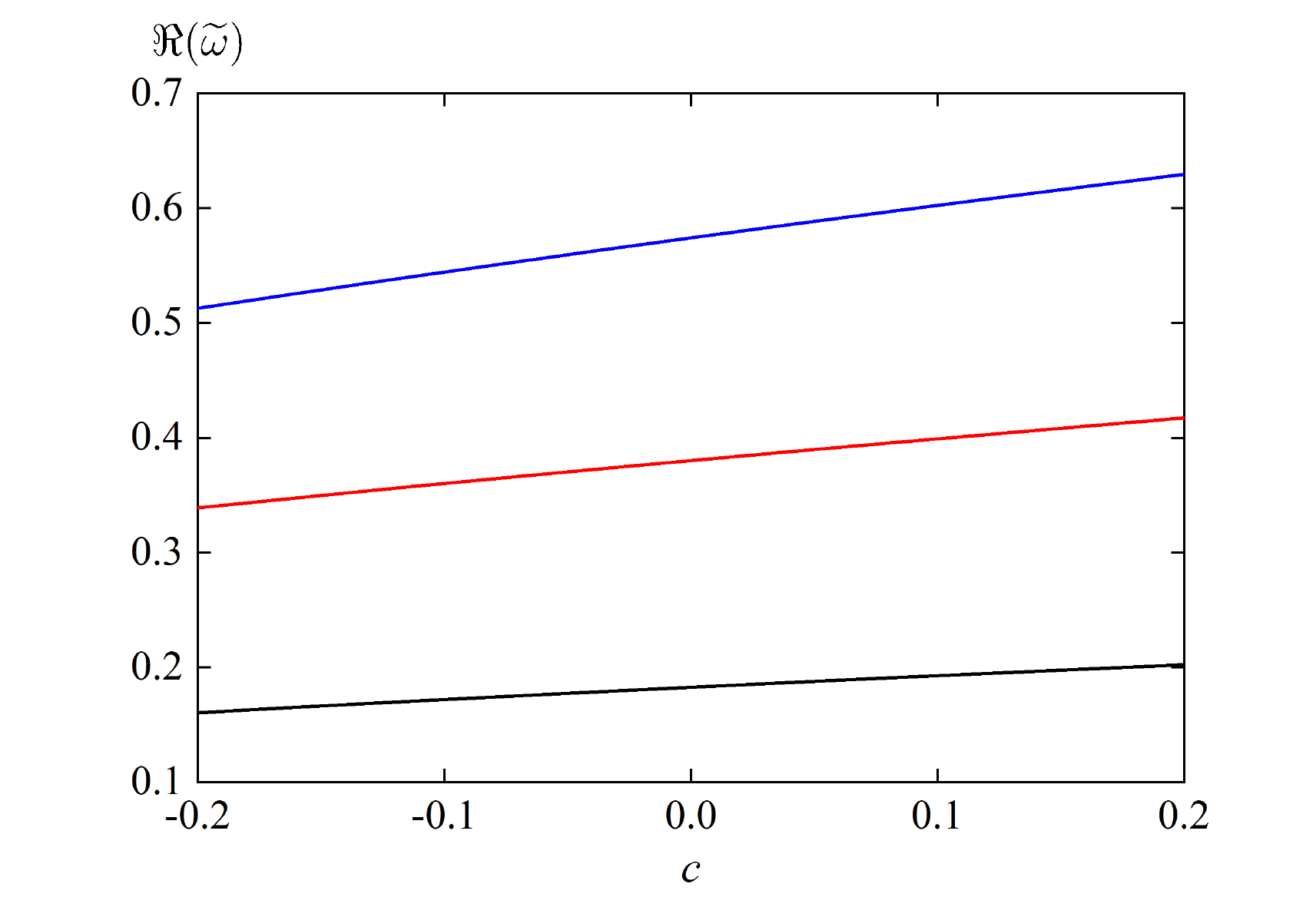}\;\;\includegraphics[clip=false,width=0.5\textwidth]{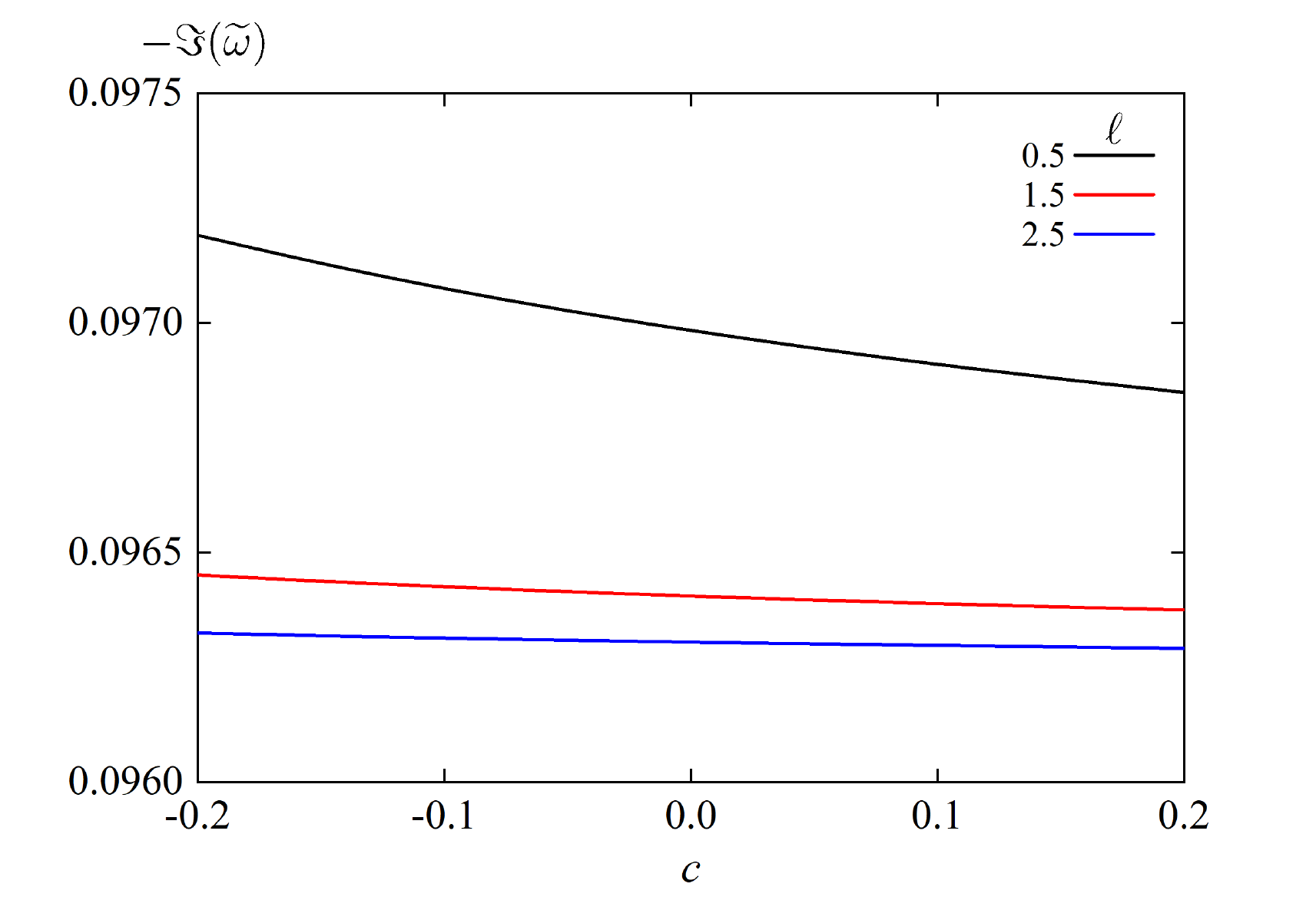}
\end{center}
\caption{\label{DiracQNMs_Bumblebee} (color online). Real (left) and imaginary (right) parts of fundamental QNMs for Dirac fields in terms of the bumblebee parameter $c$, for $\kappa=1$ (black), $\kappa=2$ (red), and $\kappa=3$ (blue), with fixed $\eta^2=0$.}
\end{figure*}

\begin{figure*}
\begin{center}
\includegraphics[clip=false,width=0.5\textwidth]{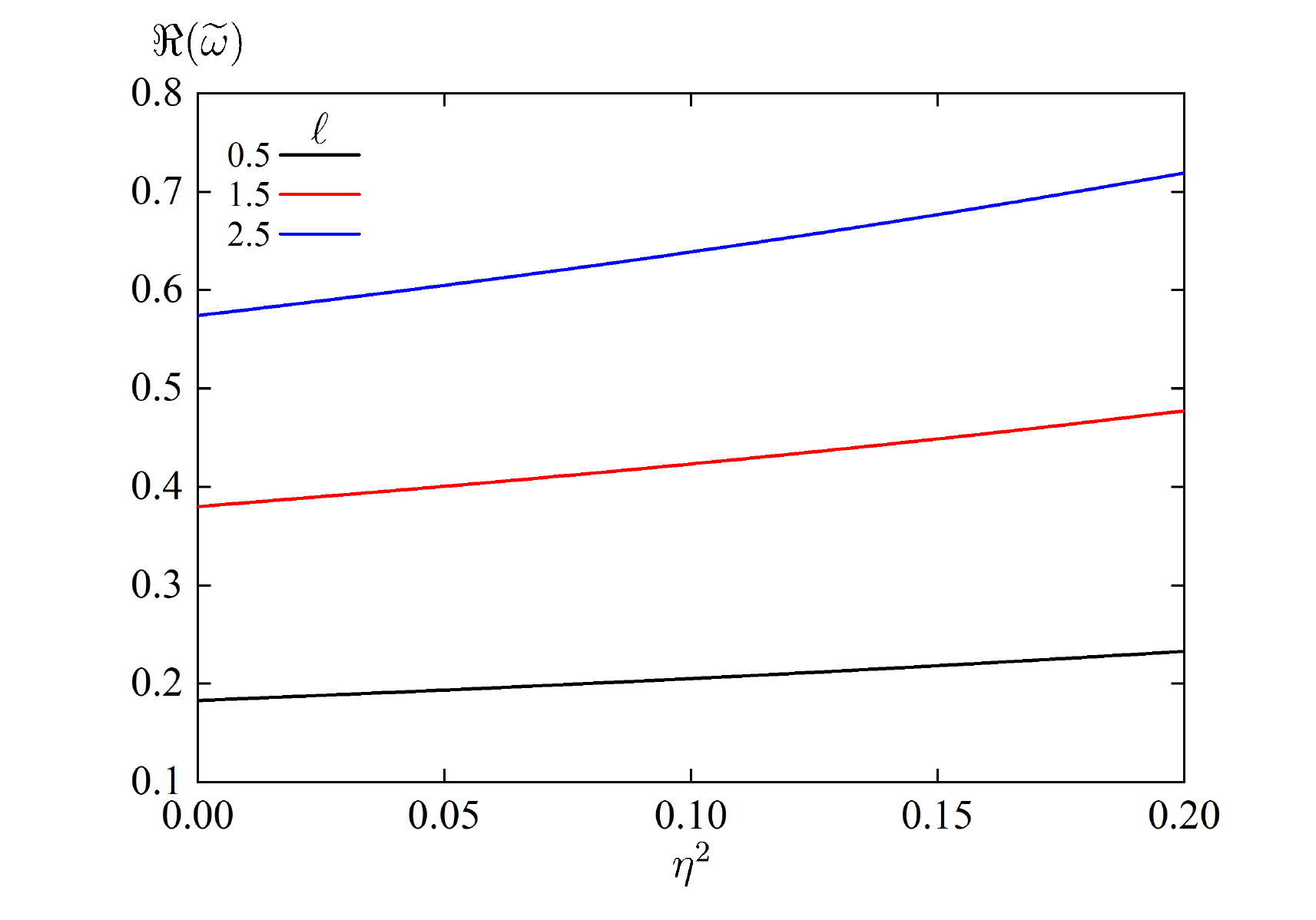}\;\;\includegraphics[clip=false,width=0.5\textwidth]{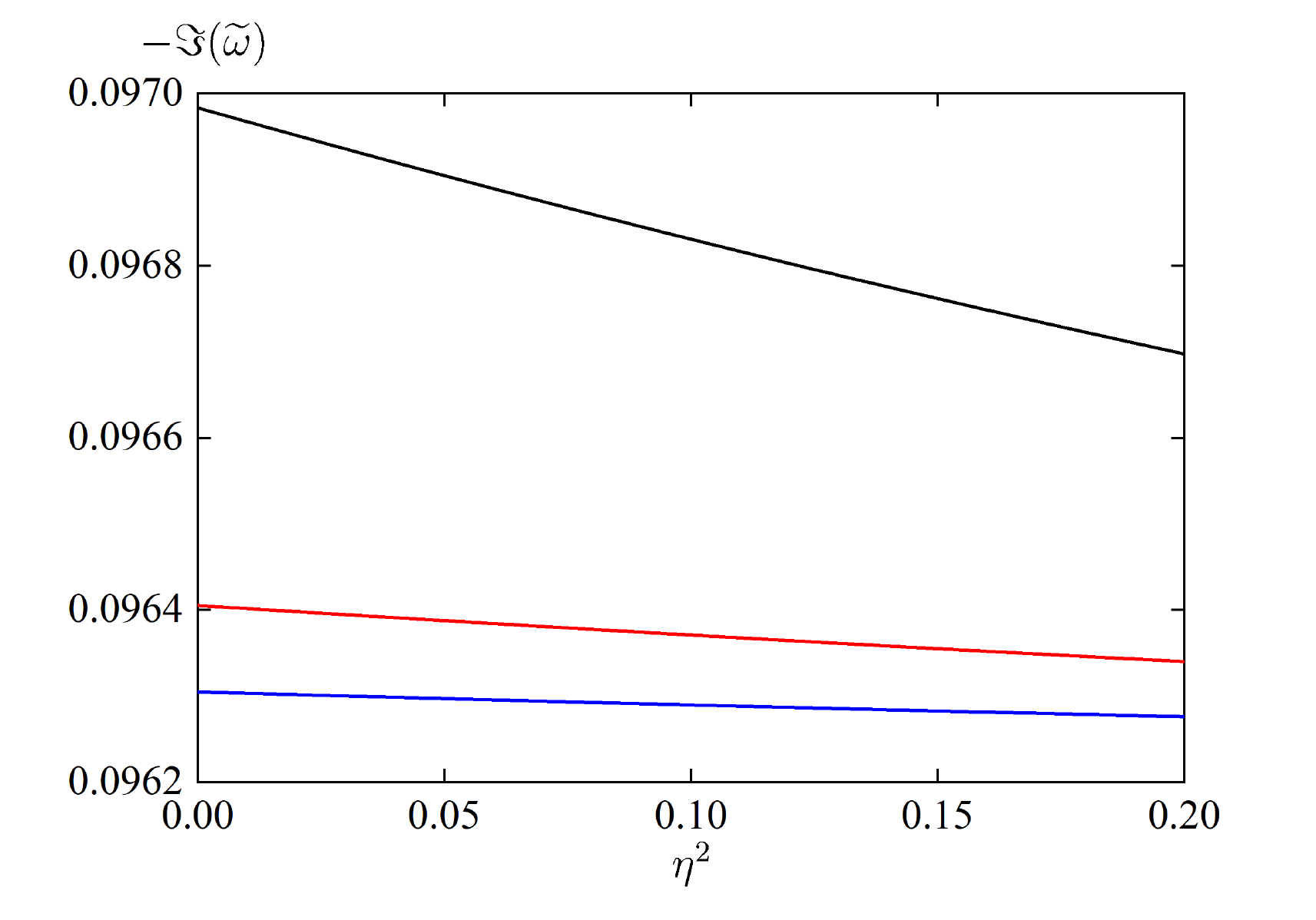}
\end{center}
\caption{\label{DiracQNMs_monopole} (color online). Real (left) and imaginary (right) parts of fundamental QNMs for Dirac fields in terms of the monopole parameter $\eta^2$, for $\kappa=1$ (black), $\kappa=2$ (red), and $\kappa=3$ (blue), with fixed $c=0$.}
\end{figure*}

Next we move to investigate the Dirac QNMs, and the corresponding results are shown in Figs.~\ref{DiracQNMs_Bumblebee} and~\ref{DiracQNMs_monopole}. In these plots, we illustrate both the real and the imaginary parts of fundamental Dirac QNMs by varying the bumblebee parameter $c$ and the monopole parameter $\eta^2$ respectively, for $\kappa=1$ (black), $\kappa=2$ (red) and $\kappa=3$ (blue). Similar to the scalar field, one observes, as $c$ ($\eta^2$) increases, that the real part of Dirac quasinormal frequencies increases but the magnitude of the imaginary part decreases. This result verifies the observation we made for the scalar case, $i.e.$ by using $\tilde{\omega}$ the bumblebee parameter has the same effect on QNMs as compared to the monopole parameter.

\begin{figure*}
\begin{center}
\includegraphics[clip=false,width=0.5\textwidth]{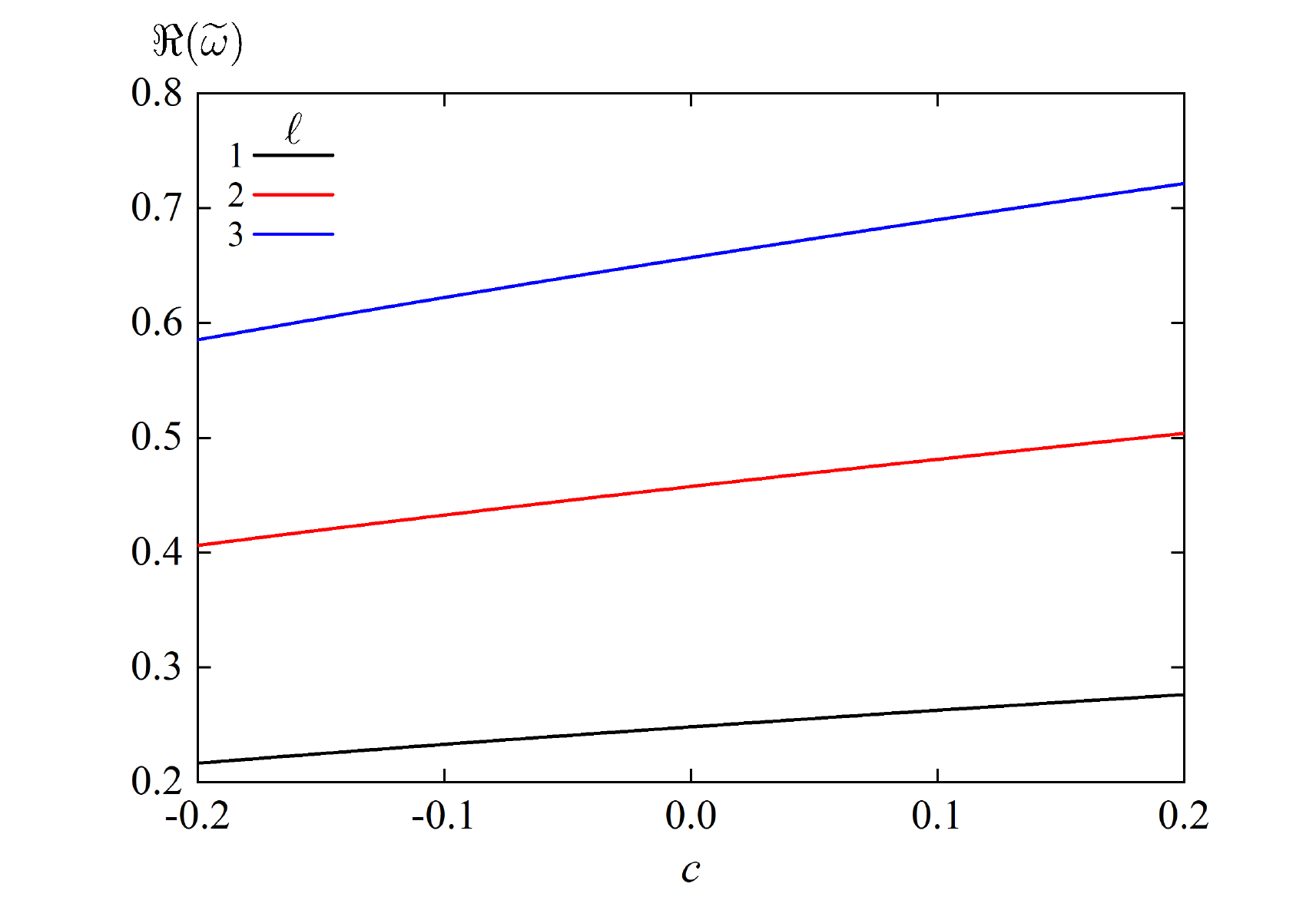}\;\;\includegraphics[clip=false,width=0.5\textwidth]{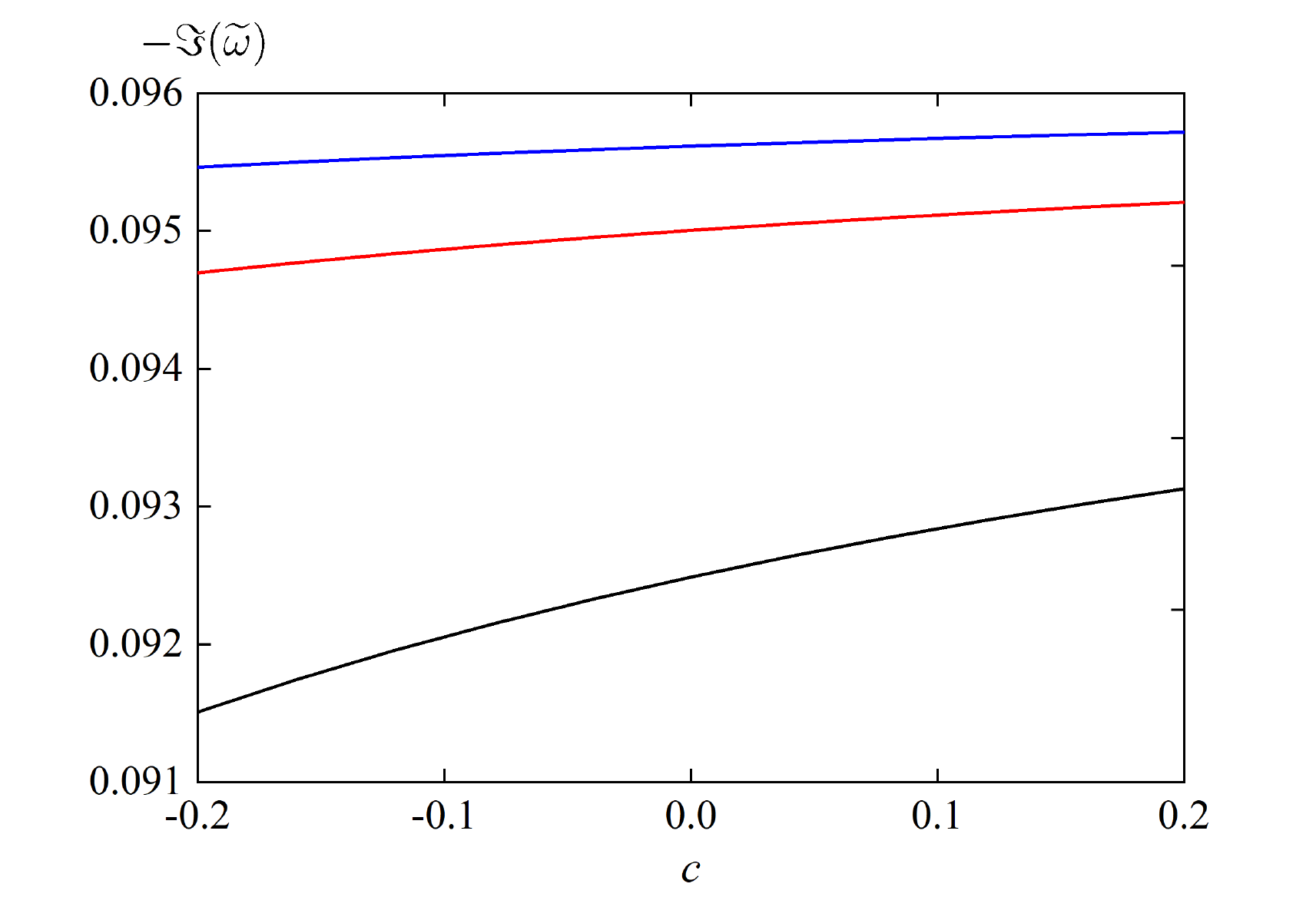}
\end{center}
\caption{\label{MaxwellQNMs_Bumblebee} (color online). Real (left) and imaginary (right) parts of fundamental QNMs for Maxwell fields in terms of the bumblebee parameter $c$, for $\ell=1$ (black), $\ell=2$ (red), and $\ell=3$ (blue), with fixed $\eta^2=0$.}
\end{figure*}

\begin{figure*}
\begin{center}
\includegraphics[clip=false,width=0.5\textwidth]{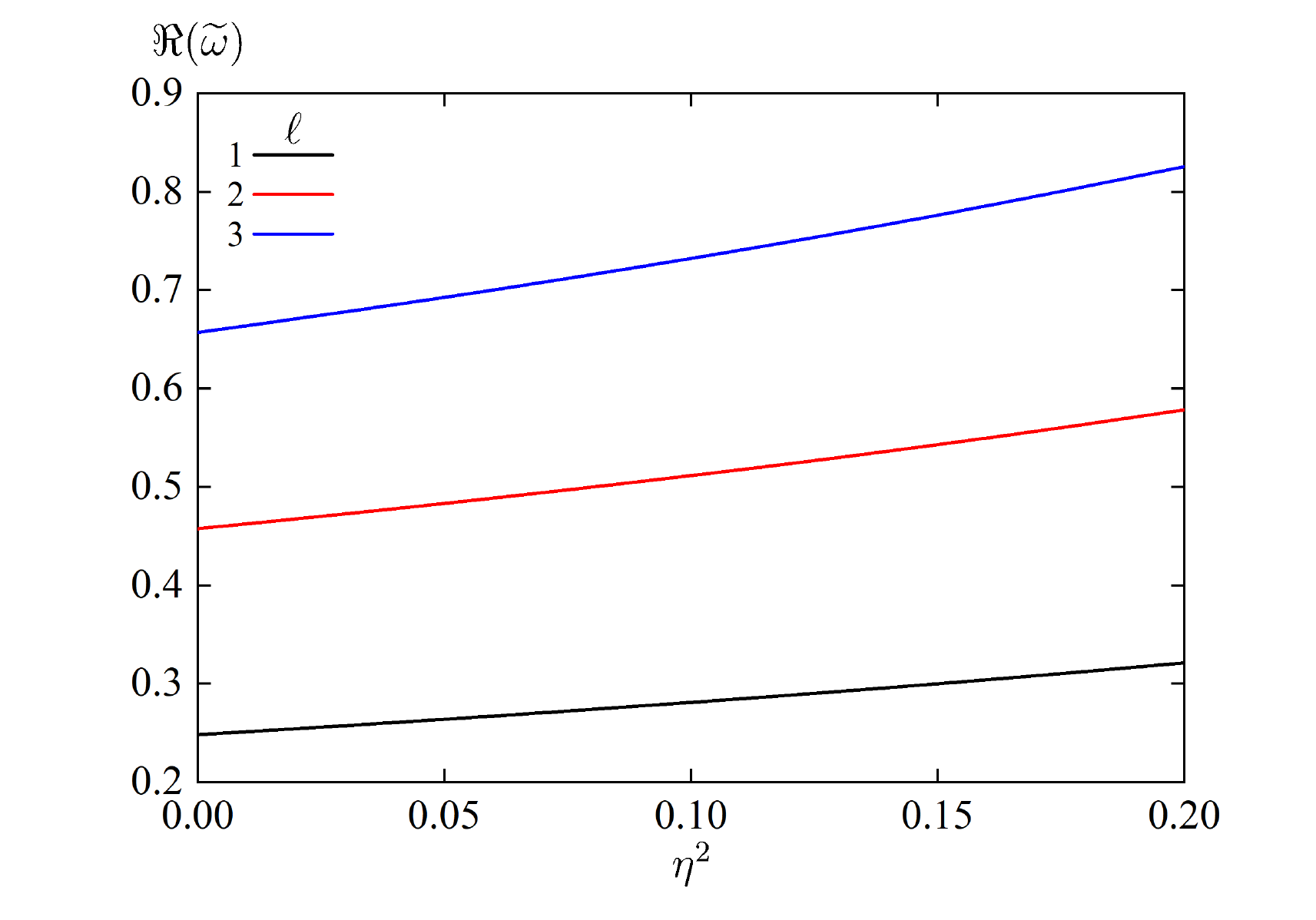}\;\;\includegraphics[clip=false,width=0.5\textwidth]{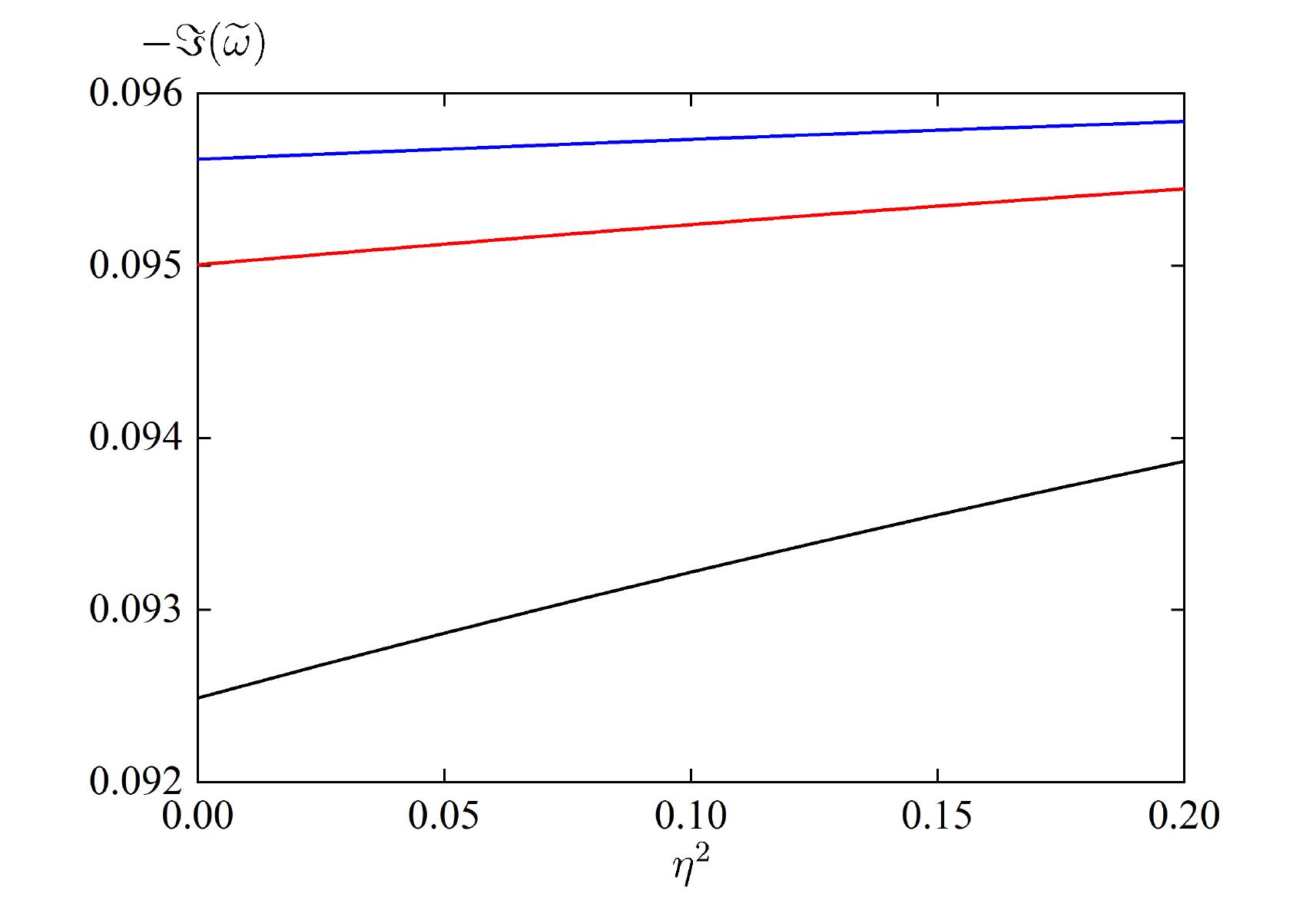}
\end{center}
\caption{\label{MaxwellQNMs_monopole} (color online). Real (left) and imaginary (right) parts of fundamental QNMs for Maxwell fields in terms of the monopole parameter $\eta^2$, for $\ell=1$ (black), $\ell=2$ (red), and $\ell=3$ (blue), with fixed $c=0$.}
\end{figure*}

As the last case we studied in this paper, the Maxwell QNMs are presented in Figs.~\ref{MaxwellQNMs_Bumblebee} and~\ref{MaxwellQNMs_monopole}. In these drawings we explore, once again, the impact of the bumblebee parameter $c$ and the monopole parameter $\eta^2$ on the fundamental Maxwell quasinormal frequencies, by taking $\ell=1$ (black), $\ell=2$ (red) and $\ell=3$ (blue) as examples. By comparing these two plots we observe, again, the fact we have announced for scalar and Dirac fields that both the bumblebee field and the monopole play the same role in determining BH quasinormal frequencies. While one observes, contrary to the previous two cases, that both the real and the magnitude of the imaginary part of Maxwell quasinormal frequencies increase as the bumblebee parameter (the monopole parameter) increases.


\begin{figure*}
\begin{center}
\includegraphics[clip=false,width=0.5\textwidth]{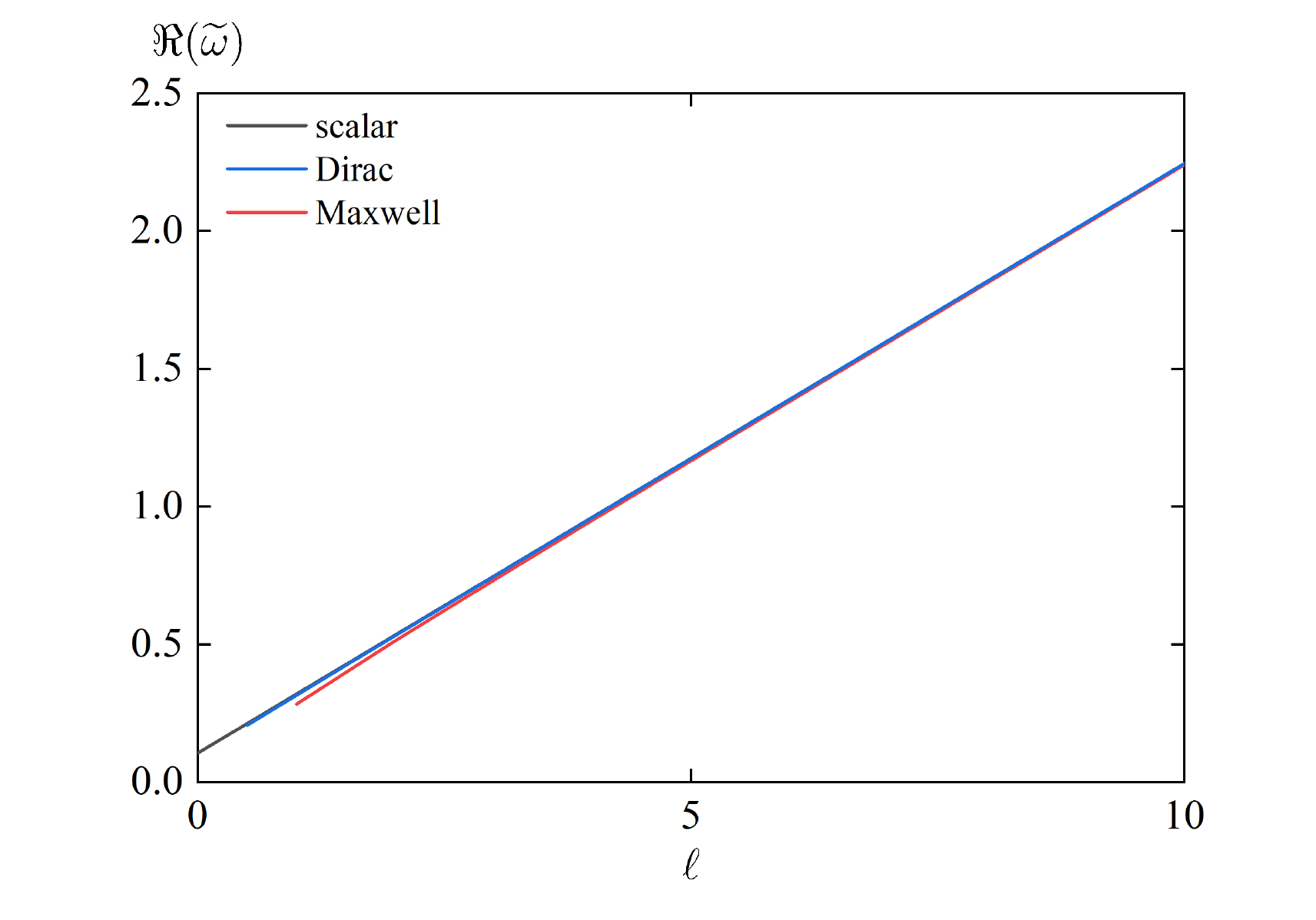}\;\;\includegraphics[clip=false,width=0.5\textwidth]{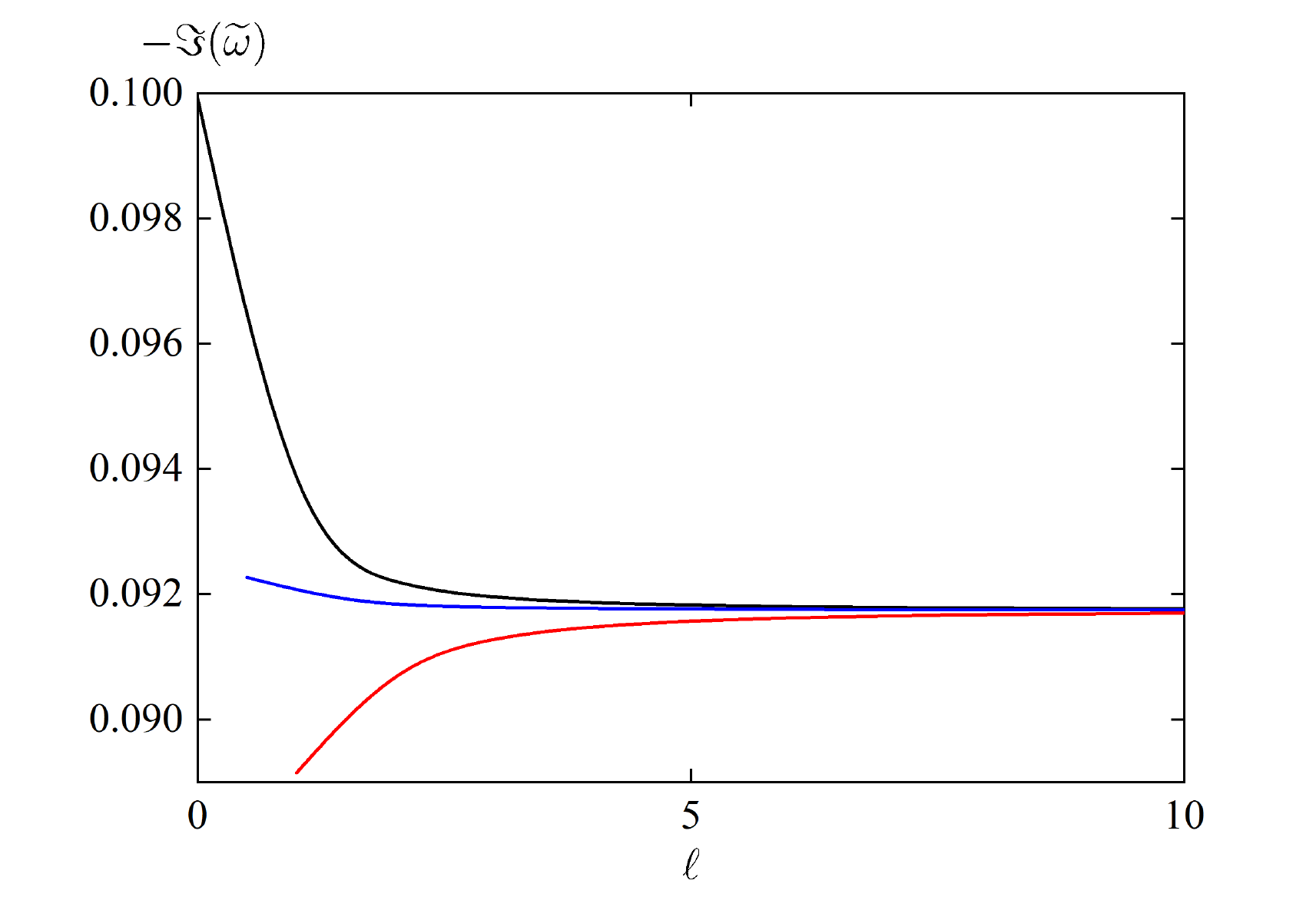}
\end{center}
\caption{\label{QNMs_ell} (color online). Real (left) and imaginary (right) parts of fundamental QNMs for scalar (black), Dirac (blue) and Maxwell (red) fields in terms of the multipole number $\ell$, with fixed $c=0.1$ and $\eta^2=0.1$.}
\end{figure*}

\begin{figure*}
\begin{center}
\includegraphics[clip=false,width=0.5\textwidth]{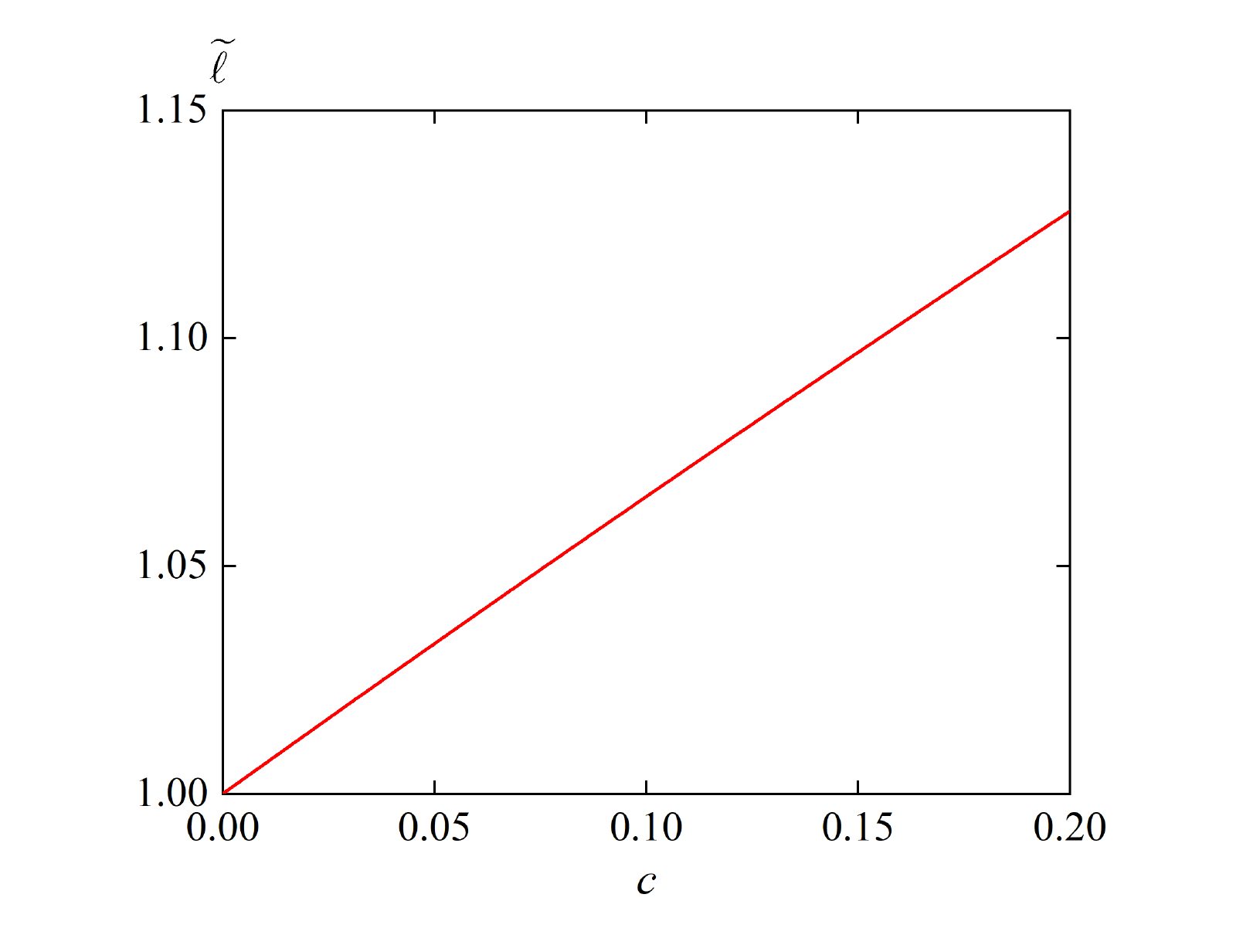}\;\;\includegraphics[clip=false,width=0.5\textwidth]{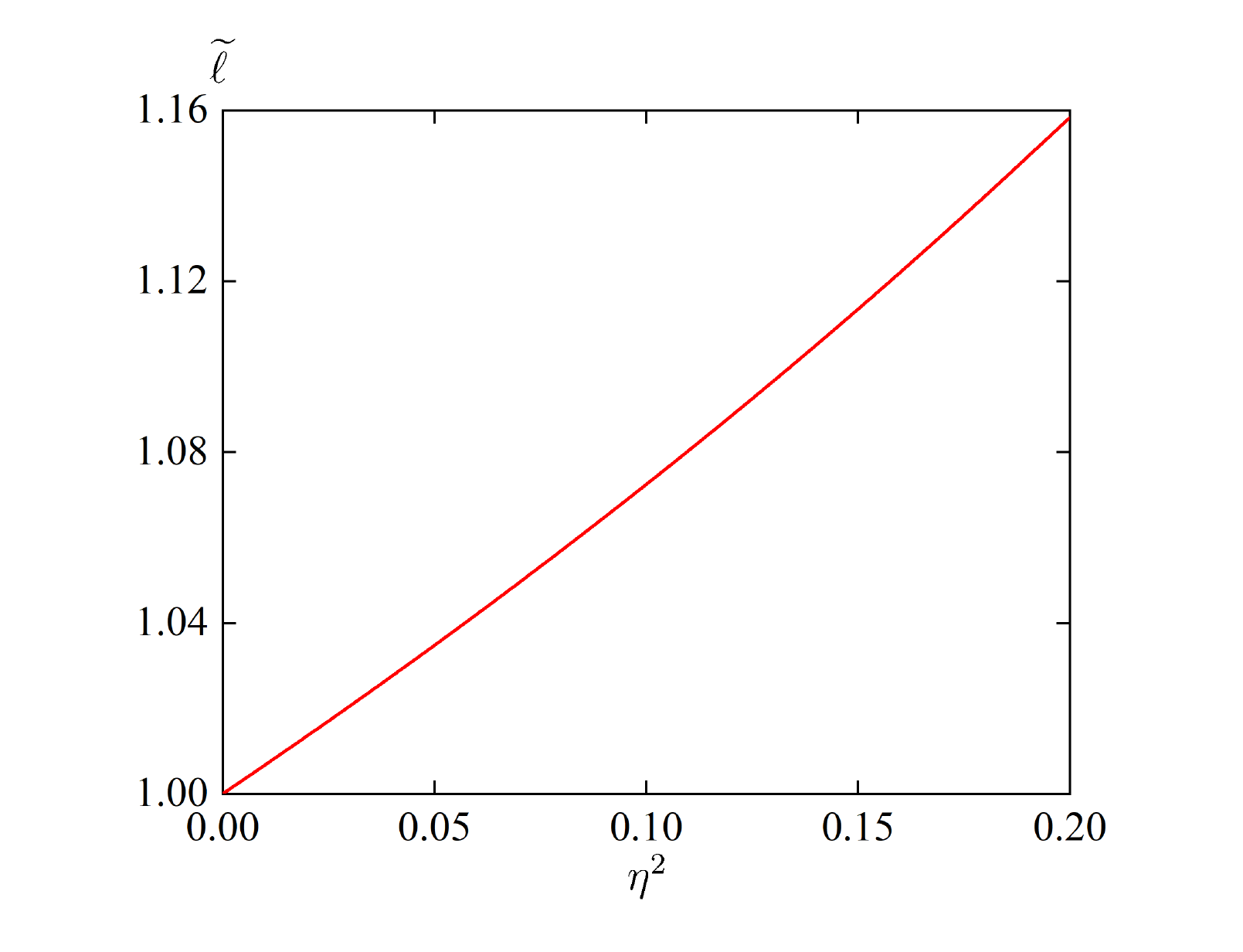}
\end{center}
\caption{\label{ellt} (color online). Dependence of the modified multipole number $\tilde{\ell}$ on the bumblebee field $c$ (left) with fixed $\eta^2=0$, and the monopole field $\eta^2$ (right) with fixed $c=0$.}
\end{figure*}

Apart from exploring imprints of the bumblebee and the monopole fields on QNMs, we have also analyzed the impact of the multipole number $\ell$ on fundamental quasinormal frequencies for scalar (black), Dirac (blue) and Maxwell (red) fields, as shown in Fig.~\ref{QNMs_ell}. It displays that, as $\ell$ increases, the real part of QNMs for all spin fields we considered in this paper increases; while the magnitude of the imaginary part of QNMs decreases for scalar and Dirac fields but increases for Maxwell fields. These behaviors may further help us to understand the dependence of QNMs on the bumblebee and monopole fields, as we observed in Figs.~\ref{scalarQNMs_Bumblebee} --~\ref{MaxwellQNMs_monopole}. To make this point more clear, in the Appendix~\ref{app}, we have rewritten equations of motion for scalar/Dirac/Maxwell field in terms of $\tilde{\omega}$. From these equations, we unveil that both the bumblebee and the monopole parameters determine quasinormal frequency $\tilde{\omega}$ by shifting the multipole number from $\ell$ to $\tilde{\ell}$. As exhibited in Fig.~\ref{ellt}, the modified multipole number $\tilde{\ell}$ increases with the bumblebee parameter $c$ (the monopole parameter $\eta^2$) increases. Based on the fact that, $(i)$ the larger bumblebee parameter (monopole parameter) indicates the larger multipole number; $(ii)$ the larger multipole number implies the larger real part of quasinormal frequency of $\tilde{\omega}$ for all spin fields while the smaller (larger) magnitude of the imaginary part for scalar and Maxwell (Dirac) fields, one immediately deduce the results we presented in Figs.~\ref{scalarQNMs_Bumblebee} --~\ref{MaxwellQNMs_monopole}.

In the eikonal regime $(\ell\gg n)$ we uncover that, for all spin fields studied in this paper, QNMs are the same and, in particular, the real part of QNMs scale linearly with $\ell$ while the imaginary part approach a constant value. These behaviors may be understood by the geometric optics approximation analysis. Based on the WKB formula, by substituting effective potentials into Eq.~\eqref{WKBorder1} and by expanding the corresponding results in the large $\ell$ limit, one obtains the asymptotic quasinormal frequencies
\begin{equation}
\tilde{\omega}\sim\dfrac{1}{3\sqrt{3}}\sqrt{\dfrac{1+c}{1-\eta^2}}\left(\ell+\dfrac{1}{2}\right)+\dfrac{i}{3\sqrt{3}}\left(n+\dfrac{1}{2}\right)+\mathcal{O}\left(\dfrac{1}{\ell}\right)\;,\label{largeell}
\end{equation}
which matches well with numeric results. From Eq.~\eqref{largeell}, a striking property appears, $i.e.$ in the eikonal limit both the bumblebee and the monopole fields \textit{only} change the real part of quasinormal frequency but have nothing to do with the imaginary part.

\section{Late time tails}
\label{secltt}
In this section, we study the late time tail behaviors of scalar, Dirac and Maxwell fields on a Schwarzschild--like BH with a global monopole in the Einstein--bumblebee theory, by using the Green's function method.

Since the late time tail behaviors are determined by the backscattering from asymptotically far regions, one may assume that both the initial data and the observer are located far away from the BH so that we can expand the equation of motion for scalar/Dirac/Maxwell field in a power series of $M/r$. Moreover, by considering the fact that the contribution to the Green's function is mainly from the low frequency part, we shall further neglect the terms $\mathcal{O}(\omega/r^2)$ in the above mentioned expansion equations. Based on these assumptions, from Eqs.~\eqref{eqS},~\eqref{eqD2} and \eqref{eqM2} with the corresponding potentials~\eqref{potentialS},~\eqref{potentialD_complex},~\eqref{potentialM} and up to the order of $1/r^2$, we get the following asymptotic equations in terms of $\tilde{\omega}$ ($\tilde{\omega}=\sqrt{1+c}\,\omega$)
\begin{equation}
\left[\dfrac{d^{2}}{dr^2}+\tilde{\omega}^2+\dfrac{4M\tilde{\omega}^2}{r}-\frac{\ell(\ell+1) (1+c)}{\bar{\eta}^2r^2}\right] \xi=0\;,
\end{equation}
for scalar and Maxwell fields, and
\begin{equation}
\left[\dfrac{d^{2}}{dr^2}+\tilde{\omega}^{2}+\dfrac{4M\tilde{\omega}^{2}-i\tilde{\omega}}{r}+\dfrac{1}{r^2}\left(\dfrac{1}{4}-\dfrac{(1+c)\lambda^2}{\bar{\eta}^2}\right)\right] \xi=0\;,
\end{equation}
for Dirac fields, and where $\xi\equiv\xi(r)=\sqrt{f(r)}R(r)$. 

Then following the standard procedures of calculating the late time tails given in~\cite{Price:1971fb,Hod:1997mt,Pan:2008xz}, we obtain the asymptotic behaviors of the Green's function at timelike infinity
\begin{align}
\begin{array}{l}
G^{c}(r_{*}, r_{*}^{\prime}; t) = \Upsilon(r_{*}, r_{*}^{\prime}) t^{-2(\rho+1)}\;,
\end{array}
\end{align}
with an integer constant $\Upsilon(r_{*}, r_{*}^{\prime})$, and where 
\begin{equation}
\rho=\sqrt{\dfrac{\ell(\ell+1)(1+c)}{1-\eta^2}+\dfrac{1}{4}}\;,\label{rhoboson}
\end{equation}
for scalar and Maxwell fields while
\begin{equation}
\rho=\left(\ell+\dfrac{1}{2}\right)\sqrt{\dfrac{1+c}{1-\eta^2}}\;,\label{rhofermion}
\end{equation}
for Dirac fields. 

It is shown, from Eqs.~\eqref{rhoboson} and~\eqref{rhofermion}, that the late time tails of bosonic fields are different from the counterparts of fermionic fields, when the bumblebee and monopole fields are present. This conclusion is contradictory with the late time tails of spin fields on Schwarzschild BHs, for which the late time tails for all spin fields are the same. Moreover, based on Eqs.~\eqref{rhoboson} and~\eqref{rhofermion}, one observes that the larger bumblebee parameter $c$ (monopole parameter $\eta^2$) leads to the larger parameter $\rho$, which not only indicates the presence of the bumblebee field (the monopole field) makes the spin fields studied herein decay faster but also implies the bumblebee and the monopole fields play the same role in determining the late time behaviors of spin fields.      

\section{Conclusion}
\label{discussion}
In this paper we have conducted a \textit{systematically} study of quasinormal frequencies and late time tails for scalar, Dirac and Maxwell fields on a Schwarzschild--like BH with a global monopole in the Einstein-bumblebee theory and, in particular, have paid attention to the influence of the bumblebee parameter $c$, the monopole parameter $\eta^2$ and the multipole number $\ell$. 

To achieve the above goal, we have performed numeric calculations to look for fundamental QNMs, by resorting to the matrix and the WKB methods. In the regime when both approaches are applicable, we found the consistent results which indicate the correctness of our data. In the numeric calculations, we take the unit of $M=1$, and measure quasinormal frequency in terms of $\sqrt{1
+c}\,M$, which is denoted by $\tilde{\omega}$ in our paper. 

In terms of $\tilde{\omega}$, we observed, for the fundamental mode, the following trends: 
\\
$\bullet$\;The bumblebee field and the monopole field have the same influence on the quasinormal spectrum.
\\
$\bullet$\;For the scalar field with $\ell=0$, both the bumblebee and monopole parameters have \textit{no} effects on QNMs.
\\
$\bullet$\;With the fixed multipole number $\ell$ ($\ell\neq0$), by increasing the bumblebee parameter $c$ (the monopole parameter $\eta^2$), the real part of QNMs increases for scalar, Dirac and Maxwell fields; while the magnitude of the imaginary part decreases for scalar and Dirac fields but increases for Maxwell fields. 
\\
$\bullet$\;With the fixed bumblebee parameter $c$ and monopole parameter $\eta^2$, as the multipole number $\ell$ increases, the real part of QNMs increases for all spin fields we considered herein; while the magnitude of the imaginary part decreases for scalar and Dirac fields but increases for Maxwell fields. 
\\
$\bullet$\;In the eikonal limit ($\ell\gg n$) and for all spin fields presented here, the real part of QNMs scale linearly with $\ell$ and approach the same value which is determined by the bumblebee parameter $c$ and the monopole parameter $\eta^2$; while the imaginary part approach a constant value which is determined by the overtone number $n$ so that has nothing to do the bumblebee and monopole parameters. The asymptotic QNMs, both for the real and for the imaginary parts, match well with the large $\ell$ limit of the first WKB formula.

In addition, we found that the late time tail depends on the bumblebee and the monopole parameters and is distinct for various spin fields. More precisely, it was shown that the late time behavior is dominated by the power law tail $t^{-2(\rho+1)}$ where, for bosonic fields (scalar and Maxwell fields), $\rho$ is given by Eq.~\eqref{rhoboson}; while for fermionic fields (Dirac fields), $\rho$ is given by Eq.~\eqref{rhofermion}. This conclusion is contradictory to the counterpart of spin fields on Schwarzschild BHs for which the late time tail for arbitrary spin fields is the same. According to the explicit expressions of $\rho$, one may affirm that the bumblebee and the monopole parameters play the same role in determining the late time behaviors for various spin fields, and the presence of the bumblebee parameter (the monopole parameter) makes the perturbation fields decay faster.

Based on the study of QNMs and late time tails, we identify that the bumblebee parameter and the monopole parameter play the same role in determining the dynamic evolution of various spin fields we explored in this paper.

To close this study, a final remark goes for the inference on the physical parameters of a Schwarzschild--like BH with a global monopole in the Einstein--bumblebee theory, through the BH spectroscopy analysis~\cite{Berti:2005ys,Berti:2016lat,Berti:2018vdi,Zhao:2022lrl,Dreyer:2003bv}. This may be achieved by dissecting the higher multipolar modes other than the dominant one, and all multipolar modes of a BH should give the same BH parameters within the measurement error range. Quite recently, the importance of overtone modes in the context of the BH spectroscopy has been verified in determining the true remnant BH patameters~\cite{Giesler:2019uxc}. By considering the low signal-to-noise of the current ground-based gravitational wave detectors~\cite{Cotesta:2022pci}, we expect to discern the impact of the bumblebee and monopole parameters on the ringdown signals by the future space-based gravitational wave detectors~\cite{Ruan:2018tsw,Lu:2019log,Moore:2014lga,Gong:2021gvw,Zhang:2021kkh}.
\bigskip

\noindent{\bf{\em Acknowledgements.}}
This work is supported by the National Natural Science Foundation of China under Grant Nos. 11705054, 11881240252, 12035005, by the Hunan Provincial Natural Science Foundation of China under Grant No. 2022JJ30367, and by the Scientific Research Fund of Hunan Provincial Education Department Grant No. 22A0039.

\bigskip
\appendix
\section{Equations of motion in terms of the frequency $\tilde{\omega}$}
\label{app}
In order to identify the influence of the bumblebee and the monopole parameters, one may rewrite equations of motions for various spin fields appeared in the main text, by using the frequency $\tilde{\omega}$ ($\tilde{\omega}=\sqrt{1+c}\,\omega$). 

For scalar fields, by multiplying $1+c$, Eq.~\eqref{eqS} becomes
\begin{equation}
\left[f\dfrac{d}{dr}\left(f\dfrac{d}{dr}\right)+\tilde{\omega}^2-f\left(\dfrac{2M}{r^3}+\dfrac{\tilde{\ell}(\tilde{\ell}+1)}{r^2}\right)\right]R=0\;,\label{appS}
\end{equation}
where $f\equiv f(r)$ with the definition given in Eq.~\eqref{metricfunc}, and
\begin{align}
\tilde{\omega}=\sqrt{1+c}\,\omega\;,\;\;\;
&\tilde{\ell}=\dfrac{1}{2}\left(-1+\dfrac{\sqrt{4\ell^2+4\ell+\alpha^2}}{\alpha}\right)\;,\label{defp}
\end{align}
with
\begin{equation}
\alpha=\sqrt{\dfrac{1-\eta^2}{1+c}}\;.\label{defalpha}
\end{equation}

For Dirac fields with real effective potentials~\eqref{potentialD_real}, by multiplying $1+c$, Eq.~\eqref{eqD2} turns into
\begin{equation}
\left[f\dfrac{d}{dr}\left(f\dfrac{d}{dr}\right)+\tilde{\omega}^2-\dfrac{f}{r^2}\tilde{\kappa}^2\mp\dfrac{\sqrt{f}}{r^2}\tilde{\kappa}\left(\dfrac{3M}{r}-1\right)\right]R=0\;,\label{appD}
\end{equation}
where $f$ and $\tilde{\omega}$ are given by Eqs.~\eqref{metricfunc} and~\eqref{defp}, respectively, and 
\begin{equation}
\tilde{\kappa}=\dfrac{\kappa}{\alpha}\;,
\end{equation}
and where $\alpha$ is given in Eq.~\eqref{defalpha}. Similar procedures may be applied to the Dirac equation with the complex potential and the corresponding equations with the frequency $\tilde{\omega}$ can be obtained straightforwardly.

For Maxwell fields, by multiplying $1+c$, Eq.~\eqref{eqM2} may be rewritten as
\begin{equation}
\left[f\dfrac{d}{dr}\left(f\dfrac{d}{dr}\right)+\tilde{\omega}^2-f\dfrac{\tilde{\ell}(\tilde{\ell}+1)}{r^2}\right]R=0\;,\label{appM}
\end{equation}
where, again, $f$ is given by Eq.~\eqref{metricfunc}, $\tilde{\omega}$ and $\tilde{\ell}$ are given in Eq.~\eqref{defp}.

The interesting fact one may unveil, from the above equations of motion for various spin fields in terms of $\tilde{\omega}$, is that these equations are the same with the corresponding equations of motion for scalar, Dirac and Maxwell fields on Schwarzschild BHs, through simply replacing the frequency $\omega$ and the multipole number $\ell$ by $\tilde{\omega}$ and $\tilde{\ell}$. This observation indicates that, for all spin fields we studied in this paper, both the bumblebee and the monopole fields alter quasinormal frequency $\tilde{\omega}$ through changing the multipole number. 

\bibliographystyle{h-physrev4}
\bibliography{BibBumblebee}

\end{document}